\documentclass[twocolumn, twocolappendix, preprint]{aastex631}
\usepackage{graphicx} % Required for inserting images
\usepackage{booktabs}
\usepackage{comment}
\usepackage{amsmath}
\usepackage{multirow}

\defcitealias{epta+inpta}{EPTA+InPTA}
\defcitealias{ng15}{NANOGrav}
\defcitealias{pptadr3}{PPTA}

\begin{document}

\title{Comparing recent PTA results on the nanohertz stochastic gravitational wave background}
% PPTA
\newcommand\swinburne{Centre for Astrophysics and Supercomputing, Swinburne University of Technology, P.O. Box 218, Hawthorn, Victoria 3122, Australia}
\newcommand\ozgrav{Australia Research Council Centre for Excellence for Gravitational Wave Discovery (OzGrav)}
\newcommand\CSIRO{Australia Telescope National Facility, CSIRO, Space and Astronomy, PO Box 76, Epping, NSW 1710, Australia}
\newcommand\MQ{Department of Physics and Astronomy and MQ Research Centre in Astronomy, Astrophysics and Astrophotonics, Macquarie University, NSW 2109, Australia}
\newcommand\monash{School of Physics and Astronomy, Monash University, VIC 3800, Australia}

% EPTA
\newcommand\apc{Universit{\'e} Paris Cit{\'e}, CNRS, Astroparticule et Cosmologie, 75013 Paris, France}
\newcommand\astron{ASTRON, Netherlands Institute for Radio Astronomy, Oude Hoogeveensedijk 4, 7991 PD, Dwingeloo, The Netherlands}
\newcommand\mpifr{Max-Planck-Institut f{\"u}r Radioastronomie, Auf dem H{\"u}gel 69, 53121 Bonn}
\newcommand\unimib{Dipartimento di Fisica ``G. Occhialini", Universit{\'a} degli Studi di Milano-Bicocca, Piazza della Scienza 3, I-20126 Milano, Italy}
\newcommand\infnunimib{INFN, Sezione di Milano-Bicocca, Piazza della Scienza 3, I-20126 Milano, Italy}
\newcommand\inafbrera{INAF - Osservatorio Astronomico di Brera, via Brera 20, I-20121 Milano, Italy}
\newcommand\tifr{Department of Astronomy and Astrophysics, Tata Institute of Fundamental Research, Homi Bhabha Road, Navy Nagar, Colaba, Mumbai 400005, India}
\newcommand\nancay{Observatoire Radioastronomique de Nan\c{c}ay, Observatoire de Paris, Universit\'e PSL, Université d'Orl\'eans, CNRS, 18330 Nan\c{c}ay, France}
\newcommand\lpcee{Laboratoire de Physique et Chimie de l'Environnement et de l'Espace, Universit\'e d'Orl\'eans / CNRS, 45071 Orl\'eans Cedex 02, France}
\newcommand\imsci{The Institute of Mathematical Sciences, C.I.T. Campus, Taramani, Chennai 600113, India}
\newcommand\iithp{Department of Physics, IIT Hyderabad, Kandi, Telangana 502284, India}
\newcommand\aei{Max Planck Institute for Gravitational Physics (Albert Einstein Institute), Am M{\"u}hlenberg 1, 14476 Potsdam, Germany}
\newcommand\ncra{National Centre for Radio Astrophysics, Pune University Campus, Pune 411007, India}
\newcommand\jbca{Jodrell Bank Centre for Astrophysics, Department of Physics and Astronomy, University of Manchester, Manchester M13 9PL, UK}
\newcommand\unibi{Fakult{\"a}t f{\"u}r Physik, Universit{\"a}t Bielefeld, Postfach 100131, 33501 Bielefeld, Germany}
\newcommand\inafoac{INAF - Osservatorio Astronomico di Cagliari, via della Scienza 5, 09047 Selargius (CA), Italy}

\author[0000-0001-5134-3925]{G.~Agazie}
\affiliation{Center for Gravitation, Cosmology and Astrophysics, Department of Physics, University of Wisconsin-Milwaukee,\\ P.O. Box 413, Milwaukee, WI 53201, USA}
\author[0000-0003-4453-3776]{J.~Antoniadis}
\affiliation{Institute of Astrophysics, FORTH, N. Plastira 100, 70013, Heraklion, Greece}
\affiliation{\mpifr}
\author[0000-0002-8935-9882]{A.~Anumarlapudi}
\affiliation{Center for Gravitation, Cosmology and Astrophysics, Department of Physics, University of Wisconsin-Milwaukee,\\ P.O. Box 413, Milwaukee, WI 53201, USA}
\author[0000-0003-0638-3340]{A.~M.~Archibald}
\affiliation{Newcastle University, NE1 7RU, UK}
\author[0000-0001-9264-8024]{P.~Arumugam}
\affiliation{Department of Physics, Indian Institute of Technology Roorkee, Roorkee-247667, India}
\author[0009-0001-3587-6622]{S.~Arumugam}
\affiliation{Department of Electrical Engineering, IIT Hyderabad, Kandi, Telangana 502284, India}
\author{Z.~Arzoumanian}
\affiliation{X-Ray Astrophysics Laboratory, NASA Goddard Space Flight Center, Code 662, Greenbelt, MD 20771, USA}
\author{J.~Askew}
\affiliation{\swinburne}
\affiliation{\ozgrav}
\author[0000-0001-7469-4250]{S.~Babak}
\affiliation{\apc}
\author[0000-0001-8640-8186]{M.~Bagchi}
\affiliation{\imsci}
\affiliation{Homi Bhabha National Institute, Training School Complex, Anushakti Nagar, Mumbai 400094, India}
% ...

% {number of names to display}{collaboration name}
\collaboration{0}{The International Pulsar Timing Array Collaboration}

\begin{abstract}
    The Australian, Chinese, European, Indian, and North American pulsar timing array (PTA) collaborations recently reported, at varying levels, evidence for the presence of a nanohertz gravitational wave background (GWB).
    Given that each PTA made different choices in modeling their data, we perform a comparison of the GWB and individual pulsar noise parameters across the results reported from the PTAs that constitute the International Pulsar Timing Array (IPTA).
    We show that despite making different modeling choices, there is no significant difference in the GWB parameters that are measured by the different PTAs, agreeing within $1\sigma$.
    The pulsar noise parameters are also consistent between different PTAs for the majority of the pulsars included in these analyses.
    We bridge the differences in modeling choices by adopting a standardized noise model for all pulsars and PTAs, finding that under this model there is a reduction in the tension in the pulsar noise parameters.
    As part of this reanalysis, we "extended" each PTA's data set by adding extra pulsars that were not timed by that PTA.
    Under these extensions, we find better constraints on the GWB amplitude and a higher signal-to-noise ratio for the Hellings and Downs correlations.
    These extensions serve as a prelude to the benefits offered by a full combination of data across all pulsars in the IPTA, i.e., the IPTA's Data Release 3, which will involve not just adding in additional pulsars, but also including data from all three PTAs where any given pulsar is timed by more than as single PTA.
\end{abstract}

\section{Introduction}

Pulsar timing arrays (PTAs) seek to detect low frequency gravitational waves (GW) by monitoring a collection of millisecond radio pulsars \citep{fb90}.
When a GW is incident on a PTA it induces shifts in the times of arrival of radio pulses.
These shifts are correlated between pairs of pulsars depending on their angular separation, known as the Helling-Downs (HD) correlations \citep{Hellings1983}.

The most likely source of low frequency GWs are supermassive black hole binaries (SMBHB), although cosmological and other more exotic sources are also possible \citep[][and references therein]{Burke-Spolaor2018-io}. 
It is expected that an ensemble of SMBHBs can generate a stochastic gravitational wave background (GWB) that could be detected first, followed by the detection of individually resolvable SMBHB sources \citep{Rosado2015}.
The spectrum of a stochastic GWB of SMBHB origin is affected by the evolution of SMBHBs and their environments \citep{Taylor2016}.
In the case of purely GW-driven evolution the pulsar timing residuals induced by the GWB follow a power law with a spectral index of $\gamma=13/3$ in the convention used in this work, although more realistic backgrounds can deviate from this expectation \citep{Phinney2001, Sesana2008, Becsy_2022}.

Currently several independent groups organized by region are operating PTAs.
The International Pulsar Timing Array \citep[IPTA,][]{Verbiest_iptadr1} is a consortium of four PTAs:
The European PTA \citep[EPTA,][]{desvignes_epta_2016}, the Indian PTA \citep[InPTA,][]{joshi_inpta_2018}, the North American Nanohertz Observatory for Gravitational waves \citep[NANOGrav,][]{ransom_nanograv_2019}, and the Australia based Parkes PTA\citep[PPTA,][]{manchester_ppta_2013}.
Additionally the Chinese PTA \citep[CPTA,][]{CPTA} and MeerKAT PTA \citep{mpta_dr1} are not yet IPTA members, but have observer status within the IPTA.

In 2020 and 2021, the EPTA, NANOGrav, and PPTA reported the discovery of a common uncorrelated red noise (CURN) process in their data, but showed no evidence for or against HD correlations \citep{epta_6psr_gwb, ng12_gwb, ppta_dr2_gwb}.
These results were supported by the analysis of the second data release of the IPTA, a combination of older data from these three PTAs, where a consistent CURN process was also found \citep{ipta_dr2, ipta_dr2_gwb}.

In \citet[][hereafter \citetalias{epta+inpta}]{epta+inpta}, \citet[][hereafter \citetalias{ng15}]{ng15}, and \citet[][hereafter \citetalias{pptadr3}]{pptadr3}, 
IPTA members presented analyses of their most recent data sets and report evidence for an HD correlated GWB with varying levels of significance.
In this work we compare the results from these three studies: describing key features and results of the published data sets in \autoref{sec:datasets}, outlining the signal and noise models in \autoref{sec:models}, comparing the properties of the GWB using published posterior samples and a new factorized analysis in \autoref{sec:compare}, and comparing individual pulsar noise properties in \autoref{sec:noise}.
Finally, in \autoref{sec:combine} we extend each PTA by adding in pulsars that are not timed by that PTA in a ``pseudo-IPTA'' combination using the factorized likelihood method.
Note that this is different from a full data combination, currently underway as part of IPTA's third data release (IPTA-DR3), where the data from every PTA will be combined for each single pulsar using a unified timing and noise model.

The CPTA also reported on an HD correlated GWB in \citet{cpta_gwb}, that appears broadly consistent with the other PTAs.
Unfortunately, we did not have access to those data and were unable to include their results in our study.

\section{Data sets and published results}
\label{sec:datasets}
\subsection{EPTA+InPTA}
\citetalias{epta+inpta} reported results from several permutations of the second data release of the EPTA \citep{epta_data}.
In this work we consider the analysis using the data \citetalias{epta+inpta} named EPTA DR2new+.
This data set contains the most recent 10.3 yr of EPTA observations for 25 pulsars.
For 10 of these pulsars, about 3.5 yr of additional observations from InPTA were combined.
The resulting combination has a total time span of about 11 yr.

The older EPTA data were excluded, because they were collected using legacy observing systems.
These observations had narrow radio bandwidths and were mostly collected at L-band (1400 MHz) only.
The lack of radio frequency coverage led to a covariance between dispersion measure variations and achromatic red noise, potentially polluting the GW information.

\citetalias{epta+inpta} reported a Bayes factor of about $60$ in favor of HD correlated GWB over CURN.
This corresponds to a false alarm probability of about $10^{-3}$ ($3\sigma$).
Assuming a power law with fixed spectral index of $13/3$, \citetalias{epta+inpta} recovered an amplitude $(2.5\pm0.7)\times 10^{-15}$ at a reference frequency of $1$ yr$^{-1}$ (median and 90\% C.I.).

\subsection{NANOGrav}
\citetalias{ng15} analyzed the NANOGrav 15 yr data set, which contains observations of 68 pulsars \citep{ng15_data}.
One pulsar was excluded from the GWB analysis, having less than 3 yr of observations.

\citetalias{ng15} reported a Bayes factor of about 200 in favor of HD correlated GWB over CURN.
This corresponds to a false alarm probability of about $10^{-3}$ or $5\times{10^{-5}}$ ($3-4\sigma$), depending on the background estimation method.
Assuming a power law with fixed spectral index of $13/3$, \citetalias{ng15} recovered an amplitude $2.4^{+0.7}_{-0.6}\times 10^{-15}$ at a reference frequency of $1$ yr$^{-1}$ (median and 90\% C.I.).

\subsection{PPTA}
\citetalias{pptadr3} analyzed the third data release of the PPTA, using observations of 32 pulsars over 18 years \citep{ppta_data}. Two pulsars were excluded from the GWB analysis because of the lack of data for one, and the presence of strong steep red noise for the other.
The full data release contains additional legacy data, which was also excluded from the GWB analysis.

\citetalias{pptadr3} reported a Bayes factor of about 1.5 in favor of HD correlated GWB over CURN and a false alarm probability of $0.02$ ($2\sigma$) based on a likelihood ratio statistic.
Assuming a power law with fixed spectral index of $13/3$, \citetalias{pptadr3} recovered an amplitude $2.0^{+0.3}_{-0.2}\times 10^{-15}$ at a reference frequency of $1$ yr$^{-1}$ (median and 68\% C.I.).

\subsection{Analysis methods}
The comparisons of published results reported in this work used samples collected by each PTA from relevant posterior distributions.
\citetalias{epta+inpta}, \citetalias{ng15}, and \citetalias{pptadr3} all used the \texttt{enterprise} software package to evaluate the likelihood and priors \citep{enterprise,enterprise_paper},
and all three PTAs used \texttt{PTMCMCSampler} to collect Markov chain Monte Carlo (MCMC) samples \citep{ptmcmc}.
All new MCMC runs for this work were conducted using the same \texttt{enterprise} and \texttt{PTMCMCSampler} software stack.

\section{Signal and noise models}
\label{sec:models}
In three series of papers the PTAs presented a variety of analyses using different signal and noise models.
Here we discuss the basic models used and point out where different PTAs do things differently.
For a comprehensive look at the noise modeling done by each PTA see \citet{epta_noise}, \citet{ng15_noise}, and \citet{ppta_noise}.

\subsection{Published analyses}
Each PTA modeled the GWB as a power law Fourier basis Gaussian process (GP) with HD correlations \citep{lentati2013, vanHaastren2009} .
The hyper parameters $\log_{10} A_\mathrm{HD}$ and $\gamma_\mathrm{HD}$ set the characteristic amplitude and spectral index of the power law, which acts as a prior on the Fourier coefficients, which describe the spectrum of the GWB.
Each PTA set the fundamental frequency of their Fourier basis as the inverse total observation time, $1/T_\mathrm{obs}$, however each PTA had a different $T_\mathrm{obs}$ resulting in different sampled frequencies.
Each PTA performed an analysis to determine the number of frequencies across which there was a significant signal in their data, and based on this used $9$, $14$ and $28$ frequencies in the GWB model for \citetalias{epta+inpta}, \citetalias{ng15}, and \citetalias{pptadr3}, respectively.
This  corresponds to frequency ranges of about $3-28$, $2-30$, and $1.7-47$ nHz, for the same.

Each PTA used a similar power law Fourier basis GP to model intrinsic pulsar red noise (RN).
Like the GWB, this noise is achromatic: it does not depend on the radio frequency of the observation.
Each pulsar had two parameters $\log_{10} A_\mathrm{RN}$ and $\gamma_\mathrm{RN}$, which characterize the power law.
The fundamental Fourier frequency for RN matched that for GWB for \citetalias{ng15}, while it varied from pulsar to pulsar for \citetalias{epta+inpta} and \citetalias{pptadr3}, corresponding to the observation time of each pulsar.
\citetalias{ng15} used $30$ frequencies to model RN and included RN for every pulsar in their array.
\citetalias{epta+inpta} performed Bayesian model selection to determine the noise  models and related optimal number of frequencies to use for each pulsar.
Pulsars with steeper RN spectra required fewer frequencies, the minimum being $10$, whereas pulsars with shallower RN spectra required more, the maximum chosen being nearly $150$.
For some pulsars there was no intrinsic RN detected, so this model was not used for those pulsars.
\citetalias{pptadr3} set the maximum RN frequency to $1/(240\, \mathrm{day})$. 
For some pulsars \citetalias{pptadr3} added an additional high-frequency achromatic noise, modeled as a power law with, maximum frequency $1/(30\, \mathrm{day})$. 
This combined model had a steep low frequency spectrum, and a shallow (but not flat) high frequency spectrum.

One of the main sources of noise in pulsar timing data are dispersion measure (DM) variations that induce a chromatic (i.e., dependent on the radio frequency $\nu$) delay in the pulse times of arrival proportional to $\nu^{-2}$.
\citetalias{epta+inpta} and \citetalias{pptadr3} used a power law Fourier basis GP to model stochastic DM variations, known as DMGP.
Again each pulsar gained two parameters $\log_{10} A_\mathrm{DM}$ and $\gamma_\mathrm{DM}$, which characterized the power law spectrum of deviations away from an initial deterministic fit.
The deterministic component included a second order Taylor series fit to the DM and a solar wind model, which took into account the relative position of the sun and pulsar when an observation was made.

\citetalias{epta+inpta} performed model selection to determine which pulsars experienced measurable DM variations and to set the number of Fourier frequencies used.
\citetalias{pptadr3} set a maximum frequency of $1/(60\,\mathrm{day})$. 

For seven and one of their pulsars, respectively, \citetalias{pptadr3} and \citetalias{epta+inpta} included an additional non-dispersive chromatic variation model, scattering noise, which modeled time delays with a $\nu^{-4}$ radio-frequency dependence.
This model used the same power law GP framework.
For each pulsar with scattering noise \citetalias{pptadr3} also added a band noise model, an additional achromatic RN GP that affects only the lowest radio frequency band, $\nu < 960$ MHz.
This band noise accounted for excess noise that arose due to assuming a fixed chromatic scaling, $\nu^{-4}$, in the scatter component. 

The main \citetalias{ng15} results we considered in this work did not use a DMGP, instead using the DMX model to account for DM variations.
DMX is a piecewise fit that determines a constant DM for each group of observations that occur near each other in time, effectively fitting a new DM for each pulsar every $\sim6$ days.
This fit was part of the pulsar timing model and was analytically marginalized as part of the GWB search.

\citetalias{pptadr3} also included additional system noise to model time-varying instrumental noise for five pulsars. This model used the achromatic power law GP framework, but applied only to observations made with a particular observing system.

\subsection{Factorized likelihood analysis}
\label{sec:models.faclik}

As described above, the three data sets were processed using distinct timing pipelines, different modeling choices, and even different Fourier bases when the same models were used.
To alleviate these procedural differences, we used the factorized likelihood approach \citep{fac_lkl} to model all the pulsars from the different data sets using a standardized noise model. Despite the possibility that this choice might be sub-optimal for some pulsars, this allowed us to make a more direct comparison between the properties of the noise and the GWB as seen by the difference PTAs.
    
In the factorized likelihood analyses, we used the NANOGrav 15 yr data set \citep{ng15_data}, the PPTA DR3 \citep{pptadr3}, and the DR2New+ data set, which is a combination of the newest EPTA data with the InPTA data \citep{epta+inpta}. For these data sets, we selected a global time span based on the longest observed pulsar in the data set, $T_{\rm span} = 18.9$~yrs.
This $T_{\rm span}$ was used to define the fundamental frequency for all models with a Fourier domain GP.
The frequencies in all Fourier bases were integer multiples $f = n / T_{\rm span}$, where $n = 1, 2, 3, \ldots, n_{\rm max}$. 
    
The standardized noise model contained four pieces: intrinsic pulsar RN, interstellar DM variations, a deterministic solar wind model, and a fixed spectral index CURN to account for the GWB.
The intrinsic RN was modeled using $n_{\rm max} = 30$ Fourier components for all pulsars.
Each pulsar had two hyperparameters characterizing the spectrum: amplitude, $\log_{10} A_{\rm RN}$, and spectral index, $\gamma_{\rm RN}$.
We modeled the DM variations in each pulsar using a $n_{\rm max} = 100$ component power law GP \citep[e.g.,][]{epta_noise}. This necessitated the removal of the DMX parameters in the pulsar timing models provided by NANOGrav, which were replaced by Taylor series expansions of DM up to a second order polynomial term, matching the DM modeling done by \citetalias{epta+inpta} and \citetalias{pptadr3}.
Each pulsar had two additional hyperparameters characterizing the DM variation spectrum: amplitude, $\log_{10} A_{\rm DM}$, and spectral index, $\gamma_{\rm DM}$.
The deterministic solar wind was modeled as described in \citet{hazboun_SW}, with the solar wind free electron density at the earth, $n_{\rm Earth}$, as the only free parameter.
The pulsar J1713+0747 has shown evidence for multiple chromatic events, causing sudden radio frequency dependent delays that relax over time, and two of these events are contained within the data sets analyzed here. We modeled these events as a deterministic exponential delay, where the amplitude, epoch, recovery time scale, and radio frequency dependence (i.e., $\nu^{-x}$) are parameters of the model \citep{ipta_dr2_gwb, epta_noise}.
Finally, the CURN was modeled as a power law process with $n_{\rm max} = 15$ Fourier components and a fixed spectral index of $\gamma_{\rm CURN} = 13/3$.
The amplitude, $\log_{10} A_{\rm CURN}$, was left as a free parameter.
After the individual pulsar analyses were completed the 
posterior distributions for the individual $\log_{10} A_{\rm CURN}$s were combined to determine a truly joint posterior for the CURN amplitude.

\section{Comparing GWB properties}
\label{sec:compare}

\begin{figure*}[t!]
    \centering
    \includegraphics[width=\textwidth]{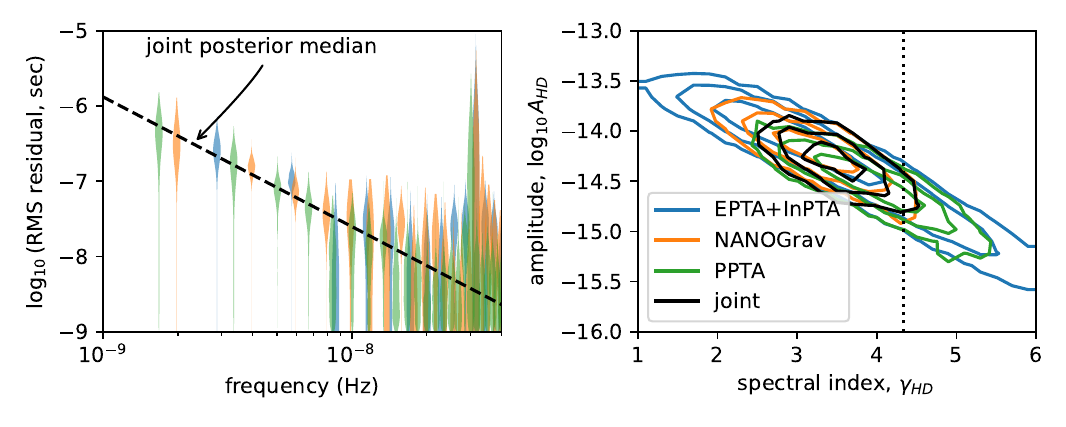}
    \caption{\textit{Left}: Free spectral posteriors for each PTA showing the measured HD correlated GWB power in several frequency bins under no spectral shape assumption. Each PTA used a different Fourier basis set by their own maximum observing time.  The dashed line shows a power law spectrum as determined by the joint 2D power law posterior median.
    \textit{Right}: 2D posterior for HD correlated power law GWB parameters.  Contours show 68, 95, and 99.7\% of the posterior mass. The vertical dotted line is at $\gamma=13/3$.
    }
    \label{fig:spec_compare}
\end{figure*}

\subsection{Comparing the published GWB measurements}
\label{sec:compare.gwb}
In previous IPTA work we compared the posterior distributions of GWB parameters as observed by different PTAs using the Mahalanobis distance, which assumes the posteriors to be multivariate Gaussian distributions \citep{ipta_dr2_gwb}.
In this work we adopted a non-Gaussian tension metric proposed by \citet{raveri2021non} to assess the tension between the posterior distributions, obtained by different PTAs.
To compute the metric, one first constructs a difference distribution by drawing a sample from each posterior distribution and taking the difference between the two, resulting in a single point of the difference distribution.
This process is repeated until the difference distribution is sufficiently mapped. Thereafter, one determines the probability that the difference is zero by integration.
The tension metric is then the Gaussian-equivalent ``$\sigma$'' value corresponding to the zero-difference probability.
We performed a detailed comparison of all GWB and noise parameters using a modified version of the \texttt{tensiometer}\footnote{\url{https://github.com/mraveri/tensiometer}} package, some of which is presented here.
The full comparison including visualizations for all GWB and noise parameters is available as supplementary material\footnote{\url{https://gitlab.com/IPTA/3pplus\_comparison\_results}}.

In \citetalias{epta+inpta}, \citetalias{ng15}, and \citetalias{pptadr3}
each PTA searched for an HD correlated GWB with a power law spectrum described by a spectral slope, $\gamma$, and characteristic amplitude, $\log_{10} A$.
The 2D marginal posterior for these parameters from each PTA is shown in the right panel of \autoref{fig:spec_compare}.
To determine the joint posterior, we computed the simple intersection of the three individual PTA posterior distributions, by multiplying them and renormalizing.
We did not weight the individual distributions in any way during this process.
This is not, strictly speaking, fair as the posteriors are not independent, some of the pulsars were observed by multiple PTAs.
However, this simplistic combination shows that combining the output of the three experiments can provide an improvement in parameter estimation.

The left panel of \autoref{fig:spec_compare} shows the free spectral posteriors from each PTA.
The individual violins show the measured HD correlated GWB amplitude at the corresponding frequency.
Additionally, the joint power law GWB posterior spectrum is plotted using the median parameters, showing it to be in good agreement with the individual free spectral results.

\begin{figure*}[t!]
    \centering
    \includegraphics[width=\textwidth]{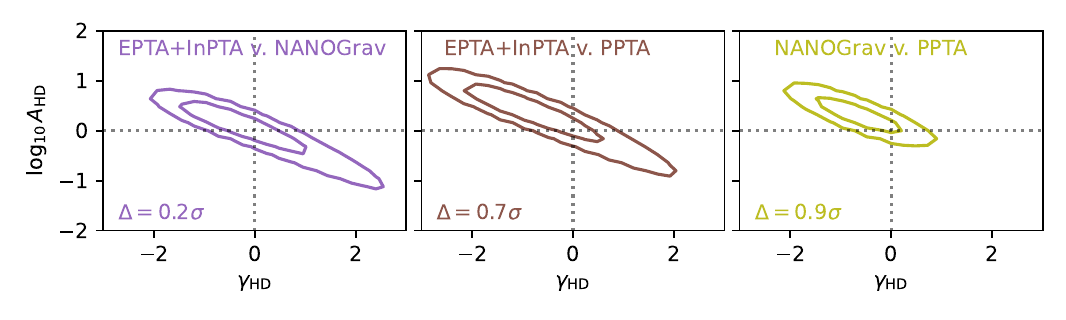}
    \caption{Difference distributions for GWB parameters between pairs of PTAs as computed by \texttt{tensiometer}.  The contours show 68 and 95\% of the distribution mass.
   }
   \label{fig:tensiometric}
\end{figure*}

\autoref{fig:tensiometric} shows the difference distributions for the power law GWB parameters for each pair of PTAs as computed by \texttt{tensiometer}.
There is excellent agreement between all three pairs as the tension metric is less than $1\sigma$ in all.
Note that \citetalias{epta+inpta} had the widest posterior distribution resulting in wider difference distributions for the two combinations involving those data.
While the numerical difference between \citetalias{ng15} and \citetalias{pptadr3} is small, the tension is larger owing to the narrower posterior distributions.

\subsection{Comparing the GWB sensitivity of PTAs}\label{ssec:sens}
\label{sec:compare.sens}
A commonly used measure of GW detector performance is a frequency-dependent `sensitivity curve'.
This metric, which estimates the smallest amplitude of a GW induced signal that a detector \textit{would} detect, is often used in the GW community to assess detector performance \citep[see][and references therein]{Moore_2015,Hazboun_2019,Kaiser_2021}. The \texttt{hasasia} \citep{hasasia} package offers a means to efficiently compute such curves for PTAs.
Specifically, the sensitivity curves we compare here are the sensitivity to interpulsar cross-correlations induced in the PTA by a GWB.
As input, \texttt{hasasia} uses the original time of arrival data and the median noise parameters for all noise processes, including the GWB auto-correlations which act as noise when trying to detect the cross-correlations.

In order to generate sensitivity curves for EPTA+InPTA and PPTA, we made a few modifications to \texttt{hasasia}. This is because \texttt{hasasia} accounts for white noise and achromatic RN only.
For analyses like \citetalias{ng15}, which modeled DM variations using DMX (which appears in the timing model) this is sufficient, but it is not for analyses that use DMGP, like \citetalias{epta+inpta} and \citetalias{pptadr3}.
The system and band noise models used by \citetalias{pptadr3} must also be accounted for\footnote{The PPTA also included a model for the variable DM delay from the solar wind, which has been left out in this analysis for simplicity, and produces the small `bump' at $\sim$2.76~nHz in the PPTA sensitivity curve.}.

\begin{figure}[!t]
    \centering
    \includegraphics[width=\linewidth]{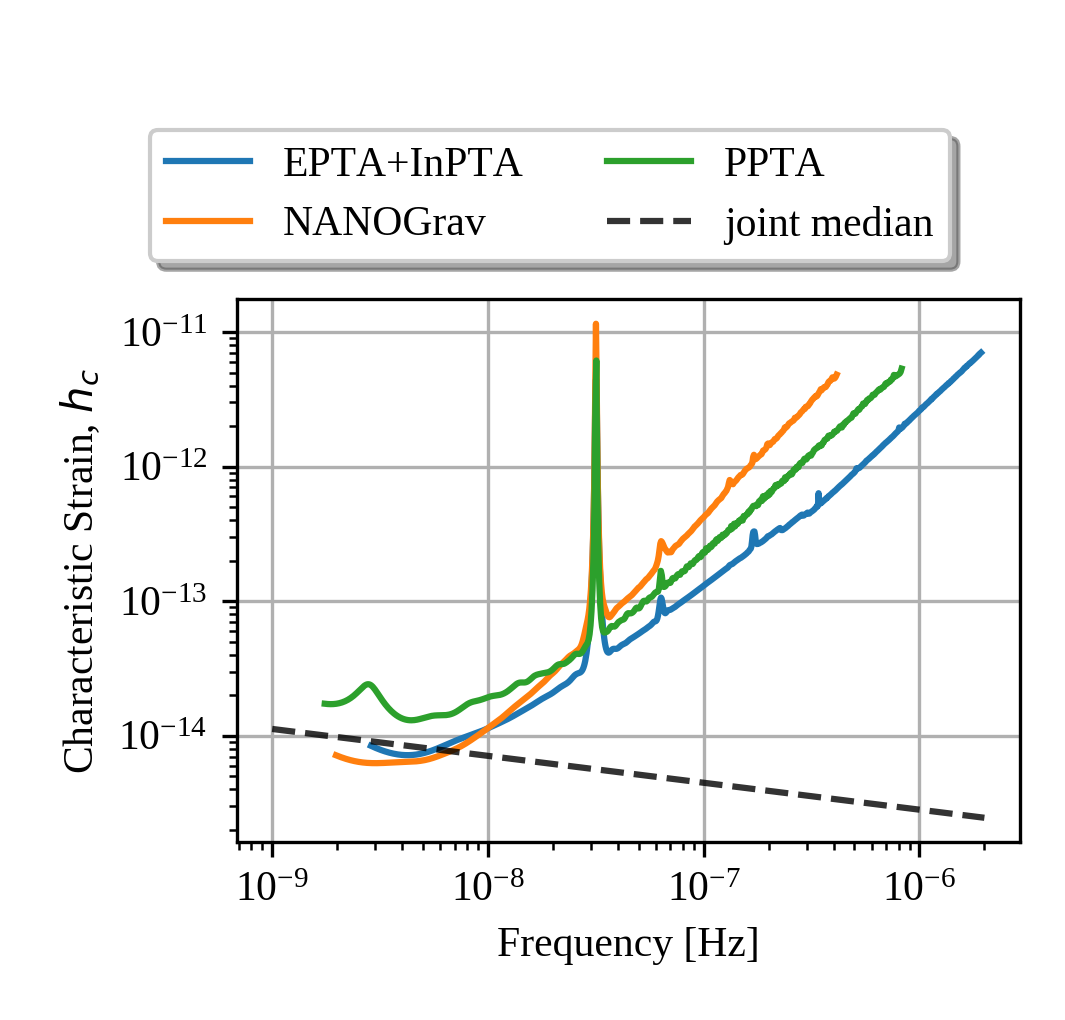}
    \caption{Estimated sensitivity to the characteristic strain induced by a GWB as a function of GW frequency. The dashed line shows a power law spectrum as determined by the joint 2D power law posterior median from the right panel of \autoref{fig:spec_compare}.}
    \label{fig:senscurves}
\end{figure}

The three resulting sensitivity curves are presented in \autoref{fig:senscurves} and show the relative sensitivity of the PTA data sets.
The general behavior of the curves follows the simple expectations based on the intrinsic properties of each data set.
The differing low frequency sensitivity supports the reported evidence from each PTA for the presence of an HD correlated GWB signal.
The NANOGrav data set shows the best low frequency sensitivity, reaching slightly lower frequencies than EPTA+InPTA due to the longer observing baseline of analyzed data. The PPTA data set which spans the longest time extends to the lowest frequencies but does not achieve the same low frequency sensitivity as the other two.
At the higher frequencies, the EPTA+InPTA and PPTA data sets are more sensitive than NANOGrav due to their higher observing cadences with observations occurring every $\sim3$ and $\sim7$ days, respectively.
NANOGrav suffers at the high frequency end due to its lower observing cadence of roughly 30 or 14\,days, depending on the pulsar.

\subsection{Comparison using standardized noise models}
\label{sec:compare.faclik}

\subsubsection{Common uncorrelated red noise amplitude}
In order to make a fair comparison of the observed GWB properties we reanalyzed each PTA's data using the standardized noise models described in \autoref{sec:models.faclik}.
In place of a full Bayesian analysis searching for HD correlated GWB, which is computationally expensive, we used the factorized likelihood approach.
Using this method individual pulsars were analyzed independently in parallel and the results combined to arrive at a posterior distribution for the amplitude of CURN, $\log_{10} A_{\rm CURN}$, assuming a fixed spectral index of $\gamma_{\rm CURN} = 13/3$.
This method did not include interpulsar cross-correlations, but acted as a good proxy to quickly determine the spectral properties of a common signal like the GWB.

\begin{figure}[t!]
    \centering
    \includegraphics[width=\columnwidth]{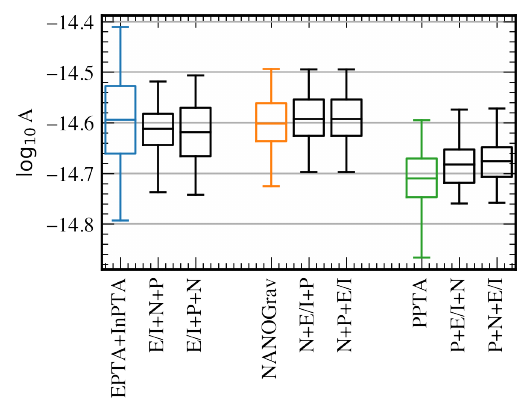}
        \caption{Amplitudes of CURN recovered using the factorized likelihood method. Extended data sets are also shown, where the pulsars of one data set are added to the another, without repeating pulsars. The boxes contain 68\% of the distribution mass, and the center line marks the median. The whiskers contain 95\% of the distribution mass.}
        \label{fig:FLAmps}
\end{figure}

We applied the standardized noise model described in \autoref{sec:models.faclik} to every pulsar with a timing baseline longer than 3 years.
Only 4 pulsars were dropped due to this time cutoff: J0614$-$3329 from NANOGrav and J0900$-$3144, J1741+1351, and J1902$-$5105 were dropped from PPTA.
Each pulsar was independently analyzed and the posteriors for the pulsars from a given PTA were combined, resulting in the CURN posteriors shown in color in \autoref{fig:FLAmps}.  
There is broad agreement between the PTAs, and these new results agree well with the fixed spectral index GWB amplitude reported by the PTAs and stated in \autoref{sec:datasets}.
The black boxes in \autoref{fig:FLAmps} are based on extending the individual PTA data sets and will be discussed in \autoref{sec:combine.faclik}.

\subsubsection{Dropout factors}
The factorized likelihood method can also be used to determine a pulsar's level of support, or lack thereof, for the CURN seen by the rest of the array \citep{ng11_cw, fac_lkl, ng12_gwb}.
This is done by using leave-one-out cross validation, comparing the ratio of Bayesian evidence for a CURN seen by the entire array versus that of the array without the pulsar in question.
This calculation gives a dropout factor (DF) for each pulsar, where a DF greater than one means that the pulsar is consistent with the common signal seen by the rest of the PTA, and a DF less than one means that the pulsar is in tension with the rest of the PTA.
A DF near one means that the pulsar does not have strong support for or against the common signal.

\begin{figure*}[t!]
    \centering
    \includegraphics[width=1\textwidth]{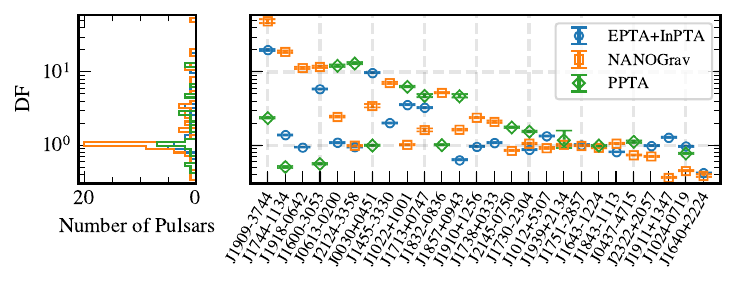}
        \caption{\textit{left}: The distribution of dropout factors (DF) for each PTA. DF is a measure of a pulsar's support for ($>1$) or against ($<1$) CURN. The majority of pulsars have a DF of one, neither supporting nor rejecting the CURN seen by the rest of the array.
        \textit{right}: DFs for pulsars that are observed by more than one PTA, allowing for data set comparison under the standardized noise model. There are serious discrepancies for many of the pulsars, indicating the importance for fully combining the data set in IPTA-DR3}
        \label{fig:FLdrop}
\end{figure*}

The left panel of \autoref{fig:FLdrop} shows the distribution of DFs for each PTA.
Many pulsars, particularly those with short observing baselines, had a DF near unity.
The right panel of \autoref{fig:FLdrop} shows the DFs for all pulsars observed by more than one PTA, with uncertainties determined by bootstrap resampling (which are very small).
There are notable discrepancies for several pulsars examined.
In some cases the differences were driven by observation time.
If a PTA had a short observing baseline for a pulsar, it should not be sensitive to low frequency effects, and it should have no support for or against CURN.
This was seen in J0030$+$0451 where PPTA have only $4$ yr of data, and the other two PTAs have more than $10$ yr.
In J2124$-$3358 the case was reversed with NANOGrav having only $3.5$ yr of data.
Three additional pulsars J1744$-$1134, J1600$-$3053, and J1857+0943 were observed with long timing baselines by each PTA, but this analysis resulted in disagreement in the DF among the PTAs.
Understanding these differences will be an important use for the forthcoming IPTA-DR3.
In some cases there was broad agreement, with pulsars like J1909-3744 and J1024-0719 having DFs greater than 1 and less than 1 for all PTAs, respectively.
Interestingly, J1713+0747 had a high dropout factor when using the PPTA data in this analysis, while in \citetalias{pptadr3} it had the lowest support for CURN.
The difference is likely attributed to noise modeling.
Our standardized noise model differs from the one employed in \citetalias{pptadr3}, where several additional components were modeled.

\subsection{Comparing interpulsar correlations}
Included in both the EPTA and NANOGrav analyses were searches for cross-correlation signals using a frequentist approach called the optimal statistic \citep{Anholm2009, Chamberlin2015}.
This approach estimates the power of a cross-correlated GWB signal, $\hat{A}^2_{\mathrm{gw}}$, by using a variance weighted least squares fit of individual pulsar pair correlated power.
This results in a signal-to-noise ratio (SNR) for a cross-correlated GWB.
The noise marginalized optimal statistic (NMOS) takes into account the posterior spread of the noise and CURN resulting in a distribution of $\hat{A}^2_{\mathrm{gw}}$ and SNRs \citep{Vigeland2018}.

Note that \cite{Romano2021} showed that the $\hat{A}^2_{\mathrm{gw}}$ recovery is biased in cases where the amplitude has a comparable effect on the residuals as the noise. This can be attributed to the fact that the optimal statistic does not account for pulsar pair covariances from a common GWB. However, the use of SNR as a detection statistic is still valid. As a result, we will focus only on the SNR calculated using the optimal statistic framework.

\begin{figure}
    \centering
    \includegraphics[width=\linewidth]{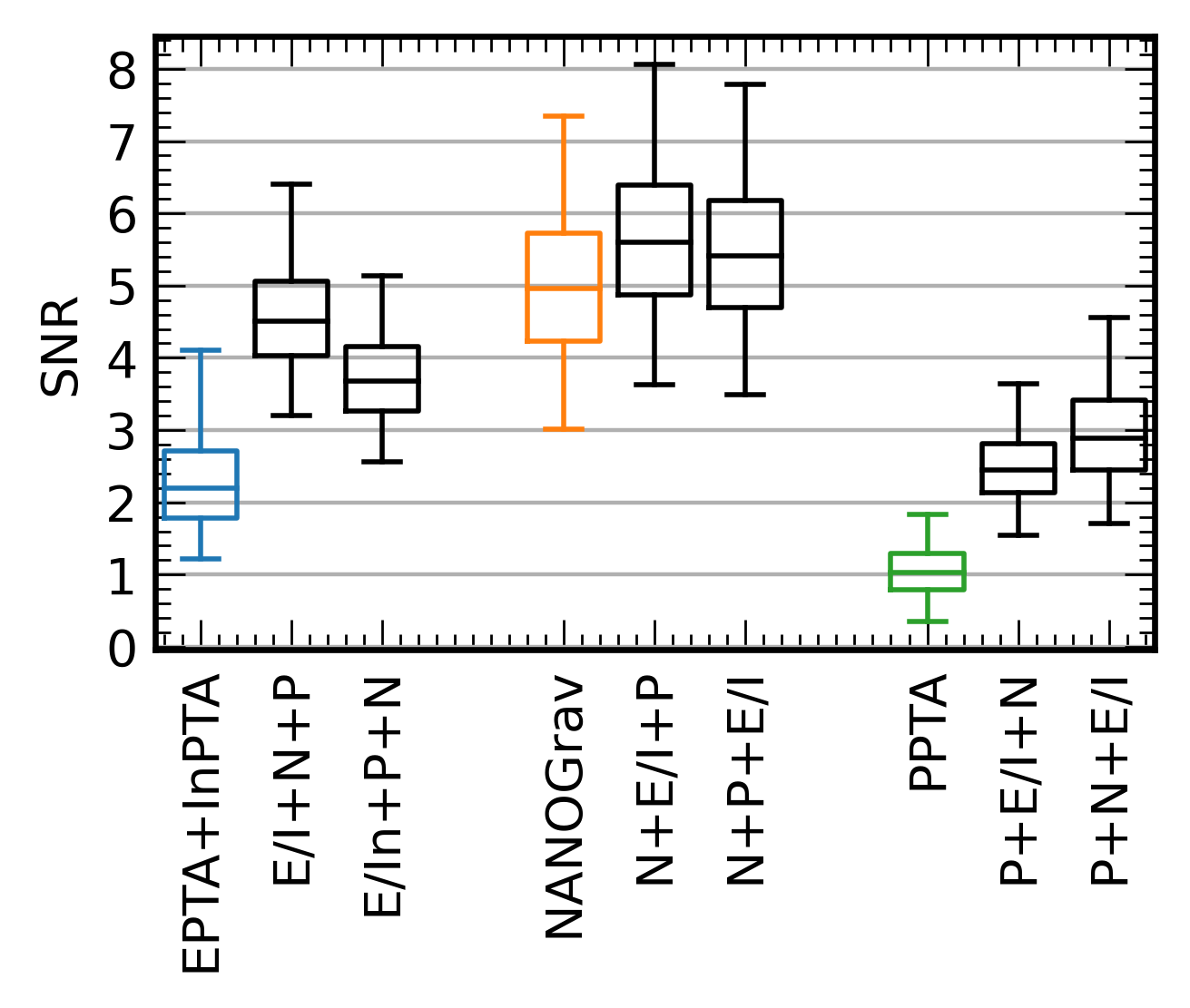}
    \caption{The noise marginalized optimal statistic distributions of signal-to-noise ratio (SNR) for each single and combination of PTAs represented as box and whiskers. The boxes contain 50\% of the distribution mass, and the center line marks the median. The whiskers contain 95\% of the distribution mass. Note that the SNR measured for the EPTA+InPTA dataset is lower than that reported in \citetalias{epta+inpta} due to our choice of Fourier basis frequencies being sub-optimal for that dataset.}
    \label{fig:nmos_combined_snr}
\end{figure}

The process of using a factorized likelihood analysis with the NMOS has been detailed within \cite{fac_lkl}.
In this case we chose to marginalize over the CURN and the intrinsic pulsar RN only, while holding the DMGP parameters fixed at their maximum likelihood values to reduce the computational complexity of the process.
The resulting SNR distributions for each PTA are shown with colors in \autoref{fig:nmos_combined_snr}. We measuree a median SNR of 2.2, 4.9, and 1.0 for \citetalias{epta+inpta}, \citetalias{ng15}, and \citetalias{pptadr3}, respectively.
The black boxes show extended data sets that will be discussed in \autoref{sec:combine.faclik}.

When comparing these NMOS results with those published by \citetalias{ng15} and \citetalias{epta+inpta}, there are some discrepancies. For \citetalias{epta+inpta} there was a decrease in SNR from around 4 to just over 2, while for \citetalias{ng15}, there was an increase in SNR from roughly 4 to 5.
Both of these differences could be attributed to the differences in pulsar noise modeling. 
Our analysis modeled each pulsar with the standardized noise model, using a Fourier basis set with a time span of just under 19 years. This is in contrast to the \citetalias{epta+inpta} published results in which they used data-driven customized noise models for each pulsar. This, combined with the fact that the time span used to define the Fourier basis here was significantly longer than the \citetalias{epta+inpta} data set may explain the overall lower SNR.
For \citetalias{ng15}, our use of DMGP differed from the fiducial DMX model and could account for our observed increase in SNR.

\section{Comparing pulsar noise properties}
\label{sec:noise}

While white noise arises during a particular observation and is partially associated with instrumental effects, other noise and signals can be astrophysical in nature and should be seen by all PTAs. 
As discussed in \autoref{sec:models} each PTA had its own methods to model noise.
Whatever the method, each PTA accounted for the two main expected noise components: achromatic RN intrinsic to each pulsar and DM variations, respectively referred as `RN' and `DM' in what follows.
Checking for consistency in these recovered noise model parameter posteriors acts as a further check of agreement between individual PTAs.
It should be noted that our standardized noise models differ from the noise modeling done by each PTA.
In particular they are simpler than those used by \citetalias{epta+inpta} and \citetalias{pptadr3} and may not be the ideal noise model for many of the pulsars considered.

There are 
$11$ pulsars that were present in all the three data sets, and
$10$, $1$, and $5$ pulsars were shared among \citetalias{epta+inpta} and \citetalias{ng15},
\citetalias{epta+inpta} and \citetalias{pptadr3}, and \citetalias{ng15} and \citetalias{pptadr3}, respectively (cf. \autoref{fig:psr_overlap}).
Here we compare their noise properties measured with each data set (1) as reported in the published posterior samples from 
noise analyses and (2) from new noise analyses that used the standardized models described in \autoref{sec:models.faclik}, but omitting the CURN component, marginalizing over solar wind electron density error from the timing model, and setting the fundamental Fourier frequency to be $1/T_{\rm{all}}$, with $T_{\rm{all}}$ the time duration between the first and the last observing epoch among all data sets.

\autoref{tab:noise_tension} shows the tension metric computed for the $27$ pulsars which were observed by multiple PTAs. For each pair of PTAs, the tensions for the RN ($\Delta \rm{RN}$) and the DM variations ($\Delta \rm{DM}$) 2D posterior distributions as reported in the published results (left) and obtained with the new analysis (right) are shown.
The large majority of cases showed consistent results among the PTAs for both RN and DM variations. For most cases, the consistency was improved with the new analysis using standardized noise models.
For some pulsars, significant tension can be explained by differences in observing time span or differences in observed radio bandwidth.
The tensions $\geq 3\sigma$ are highlighted in the table, and the main discrepancies are discussed in more detail below.

\begin{figure*}[t!]
    \centering
    \includegraphics[width=0.49\textwidth]{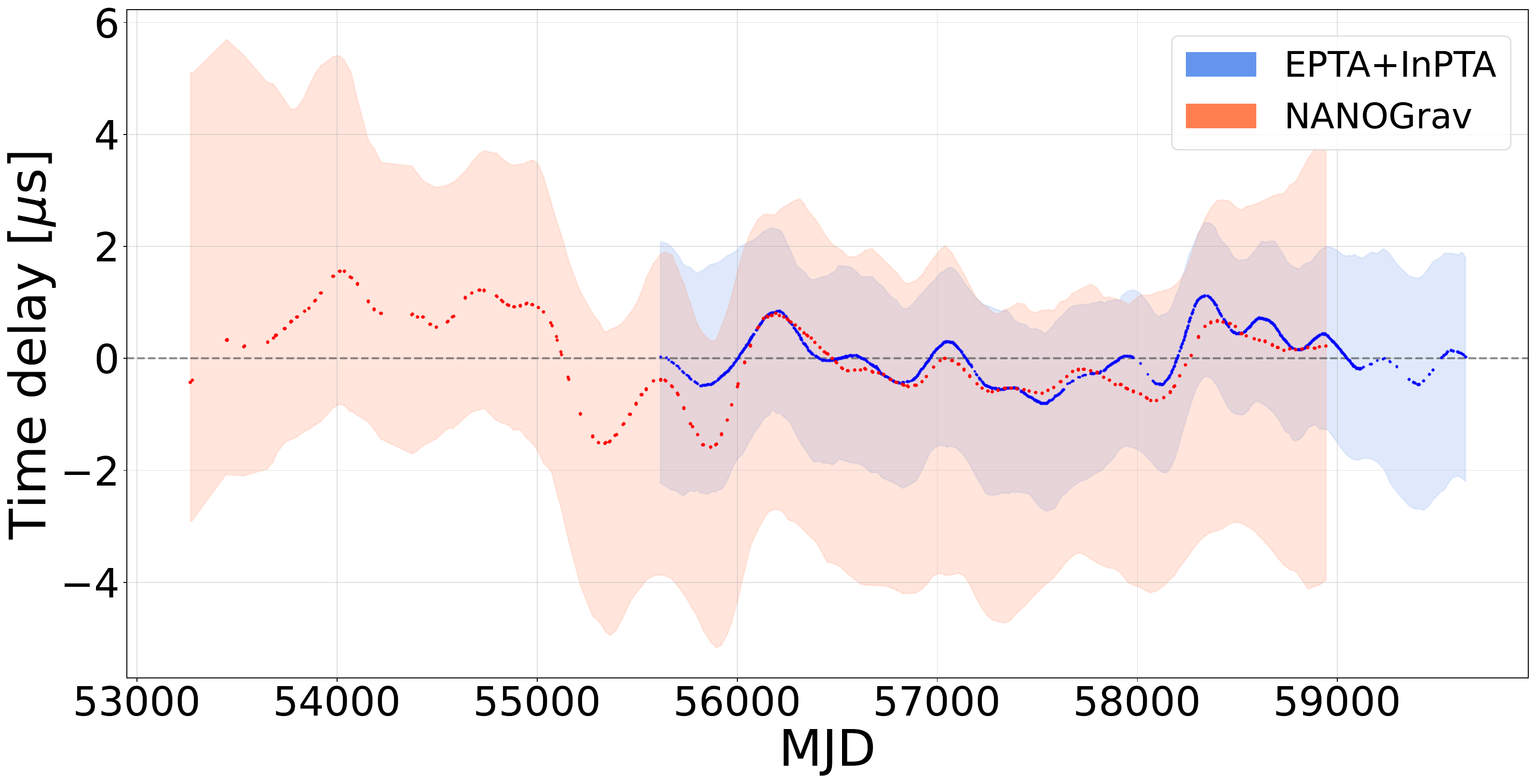} \hfil
    \includegraphics[width=0.5\textwidth]{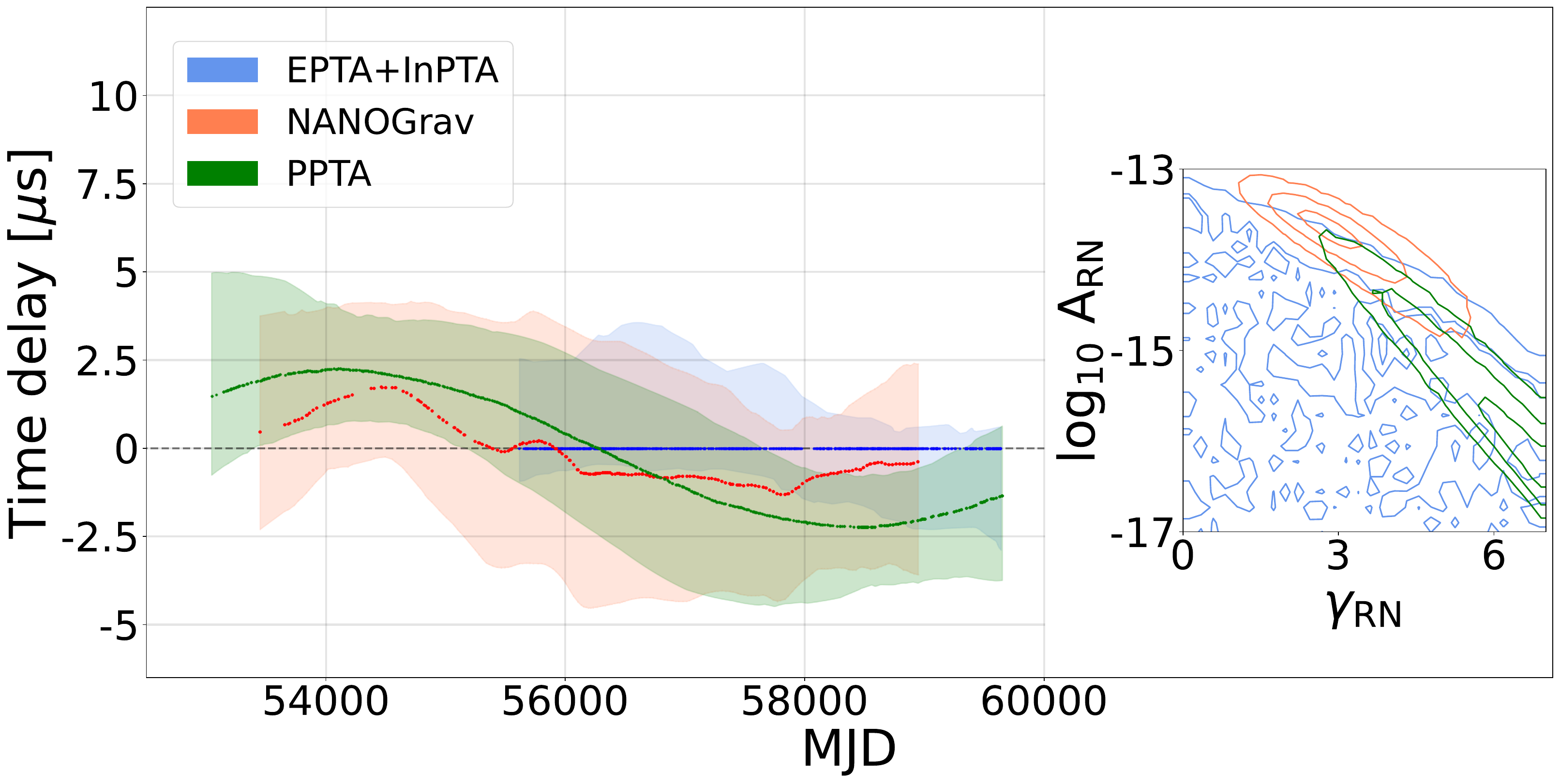}
       \caption{\textit{left:} Time-domain realizations of achromatic red noise for PSR J1012+5307 measured with \citetalias{epta+inpta} and \citetalias{ng15}, using the standardized noise models. \textit{right:} Time-domain realizations of achromatic red noise for PSR J0613$-$0200 measured with \citetalias{epta+inpta}, \citetalias{ng15} and \citetalias{pptadr3}, also using the standardized noise models. The plot at the right displays the $2$D marginalized posterior of the RN amplitude $A_{\rm{RN}}$ and spectral slope $\gamma_{\rm{RN}}$ (68, 95, 99.7\% contours). For both time-domain figures, the colored areas represent $2 \sigma$ credible intervals obtained from $300$ realizations, and the colored dots show the medians at observation epochs.
       }
       \label{fig:noiseplots}
\end{figure*}

When comparing the published results of \citetalias{epta+inpta} and \citetalias{ng15}, only the RN of PSR J1012$+$5307 has a significant tension (here $> 3.5\sigma$), which is greatly reduced when using the standardized noise models.
\autoref{fig:noiseplots} shows the time-domain realisations of the achromatic RN from the new analysis, which shows common features between the two data sets that are consistent with those reported in \citet{epta_noise}.
The tension of $2.3\sigma$ for RN of PSR J1857+0943 reflects the fact that the achromatic red noise is well constrained for \citetalias{ng15} but not for \citetalias{epta+inpta}. The tension between DM variations are measured at $1.1\sigma$ and $\geq3.5\sigma$, respectively, for PSRs J2322+2057 and J1730$-$2304. The latter is likely due to the difference in observing time span that is less than $4$ years for \citetalias{ng15} and more than $10$ years for \citetalias{epta+inpta}. A detailed explanation for the remaining pulsars with tensions $\geq 1\sigma$: PSRs J0030+0451, J0613$-$0200, J1022+1001, J1640+2224, J1713+0747, J1744$-$1134 and J2124$-$3358, is provided below. 

Comparing the published results of \citetalias{epta+inpta} and \citetalias{pptadr3}, we observed $2$ and $6$ tensions larger than $1\sigma$ for RN and DM variations, respectively.
The use of standardized noise models significantly reduced the tension in RN for both PSRs J0900$-$3144 and J1713+0747, and the tension in DM for PSRs J1022+1001, J1713+0747, J1744$-$1134, J1909-3744 and J2124$-$3358.
However, it did not improve DM consistency for PSR J0613$-$0200. Interestingly, the tension grew for the RN of PSRs J1022+1001 and J1909$-$3744 and for the DM of J1600$-$3053 and J1857+0943. Nevertheless, the time-domain realizations for the three latter were highly consistent between \citetalias{epta+inpta} and \citetalias{pptadr3}. The slight discrepancy ($< 2\sigma$) in the noise parameter posteriors for these pulsars was likely caused by the significant difference of observing time spans.

The tension between the published RN posteriors in \citetalias{ng15} and \citetalias{pptadr3} were larger than $1\sigma$ for $6$ pulsars, and significantly reduced after using standardized noise models for $5$ of them.
This clearly shows how important the choice of noise modeling is on the observed RN, which for the noise analysis implicitly included any GWB signal.
The new noise analysis resulted in more tension for PSRs J0613$-$0200 and J1744$-$1134, which are discussed below. 
The shorter observing time span of \citetalias{ng15} for PSRs J1730$-$2304 ($\sim3$ yrs vs. $>17$ yrs) and J0437$-$4715 ($\sim5$ yrs against $>15$ yrs) was very likely the cause of the large tension for these two pulsars.
For PSR J1832$-$0836, the tension of $2.4\sigma$ for DM is likely caused by a lack of cadence and radio frequency coverage for \citetalias{pptadr3}.

Let us now focus on the main remaining inconsistencies:
\begin{itemize}
    \item \textbf{PSR J0030+0451} $-$ This pulsar is known for its low ecliptic latitude, yielding significant solar wind effects that contribute to DM variations. The measured RN was consistent among the $3$ data sets ($<1\sigma$). However, the posterior distributions for the DM were fully unconstrained for \citetalias{epta+inpta} and \citetalias{pptadr3} and highly constrained to a flat power law for the \citetalias{ng15}. The lack of sensitivity was likely caused by the low radio frequency resolution for \citetalias{epta+inpta} and due to the short observing time span for \citetalias{pptadr3} ($\sim4$ yrs).
    \item \textbf{PSR J0613$-$0200} $-$ The DM variations were highly consistent between \citetalias{ng15} and \citetalias{pptadr3} ($0.7\sigma$), but \citetalias{epta+inpta} displays a flatter spectrum power law. This difference was likely caused by a longer time span of low radio frequency ($<1$ GHz) observations for the former two PTAs. The tension metric for the RN showed greater consistency between \citetalias{epta+inpta} and \citetalias{pptadr3} than for \citetalias{ng15}. However, the posterior distribution was poorly constrained for \citetalias{epta+inpta} and was constrained to differing values in the other two, as seen in \autoref{fig:noiseplots}. Despite this, the time-domain realizations in \autoref{fig:noiseplots} show a consistent long term trend for \citetalias{ng15} and \citetalias{pptadr3}.
    \item \textbf{PSR J1022$+$1001} $-$ Observations of this pulsar are more affected by the solar wind due to its low ecliptic latitude \citep{tsb2021}. It also exhibits behavior consistent with profile evolution \citep[][and references therein]{pbc20}.  Both of these factors made 
    its chromatic and achromatic noise components 
    more difficult to model than other pulsars. When using the standardized noise models, the DM variations between \citetalias{epta+inpta} and \citetalias{pptadr3} were broadly consistent ($2 \sigma$). The short observing time span for \citetalias{ng15} ($\sim5$ yr) compared with the two other ($>10$ yr) was a potential cause for the observed tension, yielding a consistent spectral slope, but a higher amplitude at $1 \ \mathrm{yr}^{-1}$. The RN was poorly constrained for both \citetalias{epta+inpta} and \citetalias{ng15}. When using standardized noise models, the tension between \citetalias{epta+inpta} and \citetalias{pptadr3} increased from $<0.1$ to $2.5 \sigma$, with the posterior distribution changing from unconstrained to constrained to a flat power law for \citetalias{pptadr3}.
    \item \textbf{PSR J1640+2224} $-$ The tension metric for the RN was only $0.2 \sigma$ between \citetalias{epta+inpta} and \citetalias{ng15}, where the former was unconstrained and the latter found a slightly constrained steep power law. However, the DM variations had a tension of $2.3 \sigma$, where \citetalias{epta+inpta} had a broader constraint for a very flat power law. As expected, we observed a better constrained posterior distribution on DM variations for \citetalias{ng15}, which had better radio frequency coverage, with low-frequency data ($< 500$ MHz) for all of the observing time span. This pulsar was not part of PPTA DR3.
    \item \textbf{PSR J1713+0747} $-$ The DM variation parameters were highly constrained and mainly consistent for the three data sets. The tension of $1.9\sigma$ between \citetalias{ng15} and \citetalias{pptadr3} reflects the difference in the amplitude that was slightly larger for the latter. All the tensions for the RN are also less than $2\sigma$, but slight differences were observed. The RN power law was flatter for \citetalias{epta+inpta} compared with the two others. Despite the very low tension ($<0.1\sigma$) between \citetalias{ng15} and \citetalias{pptadr3}, the posterior distributions contained visible differences: while the first was constrained to a single peak, the second was bimodal, with one mode consistent with \citetalias{ng15}, and the second favoring a steeper power law ($\gamma > 4.5$).
    \item \textbf{PSR J1744$-$1134} $-$ The constraints on DM variations were very consistent among the three data sets. Despite the low tension metric for the RN, we observed different behavior for each data set: unconstrained for \citetalias{epta+inpta}, steep power law broadly constrained for \citetalias{ng15} and flat power law highly constrained for \citetalias{pptadr3}.
    \item \textbf{PSR J2124$-$3358} $-$ The RN was consistent among the three data sets, being fully unconstrained for \citetalias{epta+inpta} and \citetalias{ng15}, while poorly constrained to a steep power law with \citetalias{pptadr3}. This behavior appears to be driven by the data set time spans which are $\sim4$, $\sim10$ and $>17$ years for \citetalias{ng15}, \citetalias{epta+inpta} and \citetalias{pptadr3}, respectively. However, the DM variations were fully unconstrained for \citetalias{pptadr3} and poorly constrained for the two others, favoring flat power laws.
\end{itemize}

The use of standardized noise models allowed us to show high consistency among the three data sets.
The remaining discrepancies discussed above will be studied in detail during the preparation for the full data combination of the IPTA-DR3. All plots produced for this analysis are available as supplementary material\footnote{\url{https://gitlab.com/IPTA/3pplus_comparison_results}}.

\begin{table*}[ht]
    \caption{Tension metric (in Gaussian equivalent ``$\sigma$'' units) between the data sets for noise model parameters, sorted by observing PTA. If one PTA did not observe the pulsar, ``\O''~is shown.  If one PTA did not use the relevant noise model ``$-$'' is shown. Values left of the divider are calculated using the published posteriors from individual PTAs, which make different noise modeling choices.  Values right of the divider are calculated using the new noise analysis, in which data from all PTAs are analyzed using the same noise models.  Instances with tension $\geq 3\sigma$ are in boldface.
    }
    \centering
    \label{tab:noise_tension}
    \begin{tabular}{c | c @{ $|$ } c | c @{ $|$ } c | c @{ $|$ } c | c @{ $|$ } c | c @{ $|$ } c | c @{ $|$ } c}
        \hline
        \multirow{2}*{Pulsar} & \multicolumn{6}{c|}{$\Delta$RN} & \multicolumn{6}{c}{$\Delta$DM}\\
        \cline{2-13}
        & \multicolumn{2}{c|}{E+In vs N} & \multicolumn{2}{c|}{E+In vs P} & \multicolumn{2}{c|}{N vs P} & \multicolumn{2}{c|}{E+In vs N} & \multicolumn{2}{c|}{E+In vs P} & \multicolumn{2}{c}{N vs P}\\ \hline
        J1012+5307 & $\mathbf{>3.5}$ & $0.7$ & \multicolumn{2}{c|}{\O} & \multicolumn{2}{c|}{\O} & $-$ & $0.5$ & \multicolumn{2}{c|}{\O} & \multicolumn{2}{c}{\O} \\
        J1455$-$3330 & $<0.1$ & $<0.1$ & \multicolumn{2}{c|}{\O} & \multicolumn{2}{c|}{\O} & $-$ & $0.3$ & \multicolumn{2}{c|}{\O} & \multicolumn{2}{c}{\O} \\
        J1640+2224 & $-$ & $0.2$ & \multicolumn{2}{c|}{\O} & \multicolumn{2}{c|}{\O} & $-$ & $2.3$ & \multicolumn{2}{c|}{\O} & \multicolumn{2}{c}{\O} \\
        J1738+0333 & $-$ & $0.7$ & \multicolumn{2}{c|}{\O} & \multicolumn{2}{c|}{\O} & $-$ & $0.2$ & \multicolumn{2}{c|}{\O} & \multicolumn{2}{c}{\O} \\
        J1751$-$2857 & $-$ & $<0.1$ & \multicolumn{2}{c|}{\O} & \multicolumn{2}{c|}{\O} & $-$ & $0.7$ & \multicolumn{2}{c|}{\O} & \multicolumn{2}{c}{\O} \\
        J1843$-$1113 & $-$ & $0.1$ & \multicolumn{2}{c|}{\O} & \multicolumn{2}{c|}{\O} & $-$ & $0.4$ & \multicolumn{2}{c|}{\O} & \multicolumn{2}{c}{\O} \\
        J1910+1256 & $-$ & $<0.1$ & \multicolumn{2}{c|}{\O} & \multicolumn{2}{c|}{\O} & $-$ & $0.5$ & \multicolumn{2}{c|}{\O} & \multicolumn{2}{c}{\O} \\
        J1911+1347 & $-$ & $<0.1$ & \multicolumn{2}{c|}{\O} & \multicolumn{2}{c|}{\O} & $-$ & $0.3$ & \multicolumn{2}{c|}{\O} & \multicolumn{2}{c}{\O} \\
        J1918$-$0642 & $-$ & $0.5$ & \multicolumn{2}{c|}{\O} & \multicolumn{2}{c|}{\O} & $-$ & $0.8$ & \multicolumn{2}{c|}{\O} & \multicolumn{2}{c}{\O} \\
        J2322+2057 & $-$ & $<0.1$ & \multicolumn{2}{c|}{\O} & \multicolumn{2}{c|}{\O} & $-$ & $1.1$ & \multicolumn{2}{c|}{\O} & \multicolumn{2}{c}{\O} \\
        J0900$-$3144 & \multicolumn{2}{c|}{\O} & $1.6$ & $<0.1$ & \multicolumn{2}{c|}{\O} & \multicolumn{2}{c|}{\O} & $0.7$ & $<0.1$ & \multicolumn{2}{c}{\O} \\
        J0437$-$4715 & \multicolumn{2}{c|}{\O} & \multicolumn{2}{c|}{\O} & $\mathbf{>3.5}$ & $1.1$ & \multicolumn{2}{c|}{\O} & \multicolumn{2}{c|}{\O} & $-$ & $\mathbf{>3.5}$ \\
        J1643$-$1224 & \multicolumn{2}{c|}{\O} & \multicolumn{2}{c|}{\O} & $\mathbf{>3.5}$ & $0.3$ & \multicolumn{2}{c|}{\O} & \multicolumn{2}{c|}{\O} & $-$ & $0.1$ \\
        J1832$-$0836 & \multicolumn{2}{c|}{\O} & \multicolumn{2}{c|}{\O} & $0.1$ & $0.2$ & \multicolumn{2}{c|}{\O} & \multicolumn{2}{c|}{\O} & $-$ & $2.4$ \\
        J1939+2134 (or B1937+21) & \multicolumn{2}{c|}{\O} & \multicolumn{2}{c|}{\O} & $2.9$ & $0.4$ & \multicolumn{2}{c|}{\O} & \multicolumn{2}{c|}{\O} & $-$ & $0.1$ \\
        J2145$-$0750 & \multicolumn{2}{c|}{\O} & \multicolumn{2}{c|}{\O} & $2.5$ & $<0.1$ & \multicolumn{2}{c|}{\O} & \multicolumn{2}{c|}{\O} & $-$ & $0.7$ \\
        J0030+0451 & $0.7$ & $0.5$ & $<0.1$ & $<0.1$ & $0.2$ & $0.2$ & $-$ & $1.4$ & $-$ & $<0.1$ & $-$ & $1.5$ \\
        J0613$-$0200 & $-$ & $1.8$ & $-$ & $0.5$ & $1.9$ & $2.2$ & $-$ & $\mathbf{>3.5}$ & $\mathbf{>3.5}$ & $\mathbf{>3.5}$ & $-$ & $0.7$ \\
        J1022+1001 & $0.7$ & $1.1$ & $<0.1$ & $2.5$ & $0.7$ & $0.5$ & $-$ & $\mathbf{3.3}$ & $\mathbf{>3.5}$ & $2.0$ & $-$ & $\mathbf{>3.5}$ \\
        J1024$-$0719 & $-$ & $0.1$ & $-$ & $<0.1$ & $<0.1$ & $<0.1$ & $-$ & $<0.1$ & $0.6$ & $0.2$ & $-$ & $0.4$ \\
        J1600$-$3053 & $<0.1$ & $<0.1$ & $<0.1$ & $0.3$ & $0.3$ & $0.5$ & $-$ & $0.2$ & $1.0$ & $1.2$ & $-$ & $0.4$ \\
        J1713+0747 & $1.0$ & $1.4$ & $2.3$ & $1.1$ & $2.3$ & $<0.1$ & $-$ & $0.9$ & $1.5$ & $0.3$ & $-$ & $1.9$ \\
        J1730$-$2304 & $-$ & $<0.1$ & $-$ & $<0.1$ & $0.1$ & $<0.1$ & $-$ & $\mathbf{>3.5}$ & $<0.1$ & $0.5$ & $-$ & $\mathbf{>3.5}$ \\
        J1744$-$1134 & $-$ & $1.0$ & $-$ & $0.9$ & $0.4$ & $1.7$ & $-$ & $1.1$ & $\mathbf{>3.5}$ & $1.4$ & $-$ & $0.7$ \\
        J1857+0943 (or B1855+09) & $-$ & $2.3$ & $-$ & $1.4$ & $0.5$ & $0.4$ & $-$ & $1.0$ & $0.4$ & $1.6$ & $-$ & $0.8$ \\
        J1909$-$3744 & $0.5$ & $0.4$ & $0.5$ & $1.0$ & $0.2$ & $0.2$ & $-$ & $0.7$ & $\mathbf{>3.5}$ & $0.5$ & $-$ & $0.6$ \\
        J2124$-$3358 & $-$ & $<0.1$ & $-$ & $0.5$ & $0.3$ & $0.2$ & $-$ & $1.1$ & $2.8$ & $2.4$ & $-$ & $1.2$ \\
        \hline
    \end{tabular}
\end{table*}

\section{Extending PTA data sets with additional pulsars} \label{sec:combine}

\subsection{Combined free spectral refit}
A realistic GWB spectrum will likely have more complicated features than a simple power law.
The $\gamma=13/3$ power law is the simplest model for a population of circular SMBHBs that are driven towards inspiral purely due to GW emission \citep{Phinney2001}, although realistic backgrounds will deviate from this primitive spectral model \citep{Sesana2008, Becsy_2022}. For example, three-body interactions between the SMBHB and stars in the loss cone and interactions between the SMBHB and its circumbinary gas disk will also drive the inspiral at larger SMBHB separations \citep{Sesana_2013, Burke-Spolaor2018-io}. This attenuates the GWB spectrum at lower frequencies, creating a `turnover' spectrum \citep{Sampson_2015}.
Detecting such a turnover could further resolve the `last parsec problem' \citep{milosavljevic2003long, khan2013supermassive, vasiliev2014final, ng15_astro, epta_interp}, though we note that other sources have been proposed as a source of the GWB that can also produce a turnover spectrum \citep{ng15_cosmo, epta_interp}.
This turnover spectrum can be modeled using two power laws with different spectral indices and a bend frequency where the transition occurs.

Typically, the GWB is assumed to be a Gaussian process. However, if SMBHBs are the source of the signal, it is possible that the occupation fraction of SMBHBs emitting in each frequency bin is too small and will cause a deviation from the Gaussianity assumption.
To account for this possibility, we can apply a weight factor to each frequency of the standard power law which is distributed by a Student's $t$ distribution, modeling non-Gaussian deviations from the power law spectrum. This model is known as a `$t$-process' \citep{ng15}. A localized peak in the spectrum will result in a large weight measured at that frequency.

The forthcoming IPTA-DR3 combined data set will allow for a precise search for turnovers or deviations away from a pure power law in the GWB spectrum. However, we can make a pseudo-IPTA combination using the fast spectral refit methods of \citet{Lamb_2023}, where we can estimate the recovered spectral index and amplitude by simultaneously refitting to the published free spectral posteriors of the three data sets (see \autoref{fig:spec_compare}).

We used the \texttt{ceffyl} \citep{Lamb_2023} software package to represent each posterior with highly optimised kernel density estimators (KDEs). For a given set of spectral parameters, we computed the probability density of the GWB spectral model at each frequency according to the KDE. The effective likelihood is simply the product of these probability densities over frequency bins. We confirmed that the posterior recovered when fitting a power law using this method is consistent with the joint posterior shown in \autoref{sec:compare.gwb}.

\begin{figure*}[t!]
    \centering
    \includegraphics[width=\textwidth]{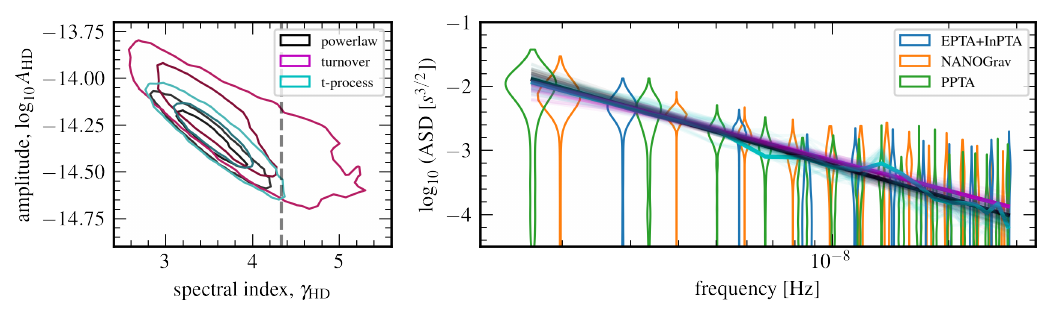}
    \caption{Simultaneously fitting GWB spectral models to the free spectra of the three data sets. \textit{Left:} 68\%- and 95\%-credible regions for the recovered GWB amplitude and spectral index for a power law, turnover, and $t$-process model. They are all consistent, however the turnover model is significantly less constrained and has more support for a $\gamma=13/3$ process. \textit{Right:} The free spectral posteriors from each data set overlaid by power laws constructed from random draws from the \texttt{ceffyl}-generated posteriors. Colors correspond to the models on the left panel.}
    \label{fig:ceffyl_compare}
\end{figure*}

The posteriors for this joint Bayesian fitting are shown in \autoref{fig:ceffyl_compare}. We recovered consistent posteriors across the turnover, $t$-process and power law models, which suggests that no spectral features beyond a power law are favored by the data.
The turnover spectrum had a broadened posterior, showing more support for the $\gamma=13/3$ power law.
However, this was because the weak constraints on the first frequency bins of each data set allowed some support for a turnover near the low frequency cutoff, and hence a steeper higher-frequency spectral index than for a pure power law. The $t$-process did not show significant support for any deviations away from the power law, except at high frequencies where the spectrum was white noise dominated.
There are some caveats to this result. As explained previously, each PTA used different methods to model their noise sources, such as DM variations. In addition, some pulsars were included in two or more of the data sets and are therefore being ``double counted'' in this pseudo-IPTA combination. 

\subsection{Extending individual PTAs using factorized likelihood}
\label{sec:combine.faclik}
The flexibility and speed of the factorized likelihood method introduced in \autoref{sec:models.faclik} provides an immediate approach to extending individual PTAs by sequentially adding pulsars observed by one PTA to those from another (being careful to only have one version of each pulsar) to achieve a pseudo-IPTA data set. We did this pseudo-IPTA construction in a piece-wise fashion, where we first chose a ``base'' PTA data set, and then added in pulsars from other data sets that are not timed by the base data set.

\begin{figure}
    \centering
    \includegraphics{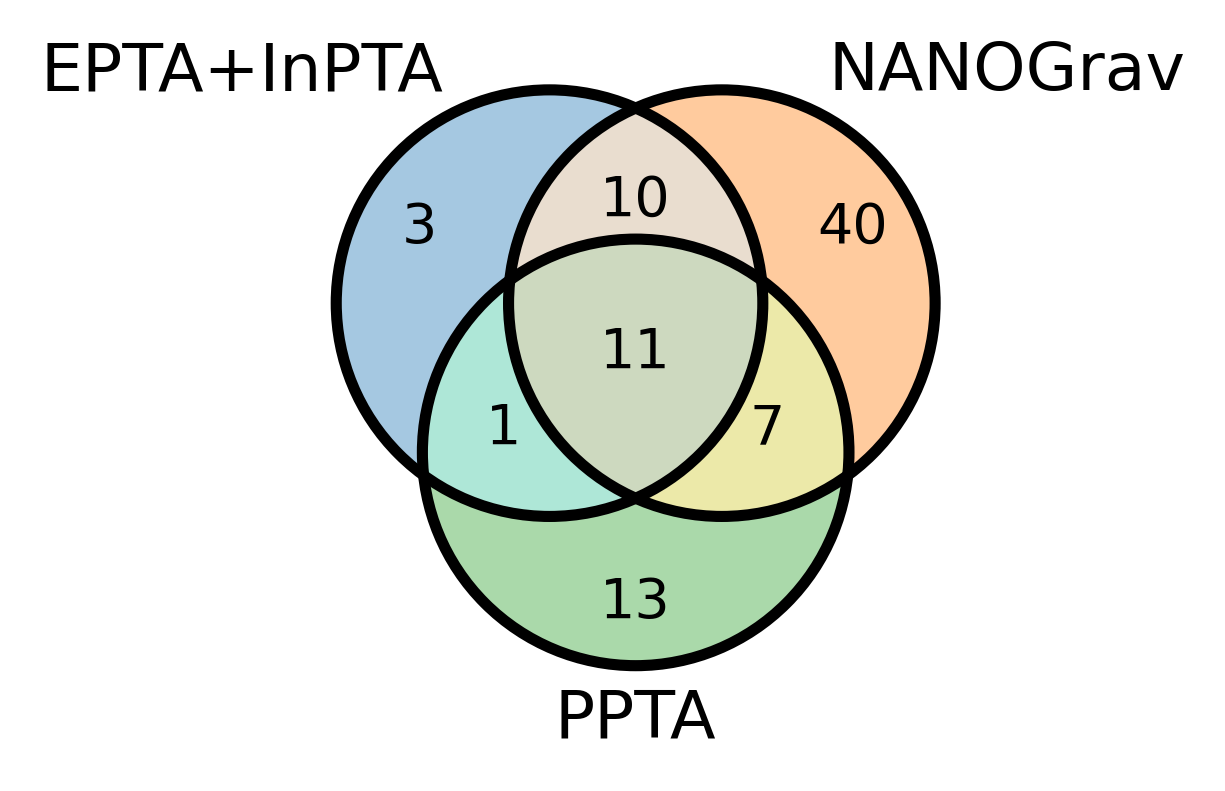}
    \caption{An unweighted Venn diagram showing the overlapping pulsars between each PTA's data sets. Note that for the factorized likelihood analyses 4 pulsars were dropped from the PPTA DR3 due to baselines for those pulsars being shorter than 3 years, but are included in this diagram for completeness.}
    \label{fig:psr_overlap}
\end{figure}

Since we only add the pulsars not already present in the data set, this addition operation is not commutative. For example, adding the new pulsars from EPTA+InPTA to NANOGrav means that the pulsars in common will use NANOGrav data. The number of overlapping pulsars for each PTA are shown in \autoref{fig:psr_overlap} where EPTA+InPTA is represented by the blue circle, NANOGrav by orange, and PPTA by green.  Combining all 3 data sets results in 6 different permutations to consider. This is similar to the approach of \citet{ipta_dr2_lite} to combine individual pulsars from different PTAs in a ``lite'' combination.

The effect of adding pulsars within the factorized likelihood analysis, producing a pseudo-IPTA data set, can be seen in the set of CURN posteriors which are shown in black in \autoref{fig:FLAmps}.
In each case, adding pulsars drawn from the other PTAs resulted in a more constrained CURN amplitude posterior than given by the original PTA alone, with a median thinning of the 68\% credible interval by 15\%. As a direct result of this, the Bayes factor for CURN over intrinsic RN alone also increased, with a median increase of 7 orders of magnitude. This is consistent with the findings of the IPTA-DR2 analysis \citep{ipta_dr2_gwb}, where improvements to parameter estimation and detection significance are observed when PTA data sets are extended. 

Starting from each of the factorized likelihood CURN posterior from the above analysis, we again used the NMOS to determine the SNR distributions for each pseudo-IPTA data set.
This method shows how adding in pulsars from other PTAs affects the significance of correlated power.

\autoref{fig:nmos_combined_snr} shows each individual PTA's NMOS SNR recovery alongside each possible pseudo-IPTA data set.
These SNR distributions show that it did not matter which PTA we started with, adding additional pulsars will always result in a higher SNR. This is consistent with scaling relations for the optimal statistic found in \cite{Siemens2013}, and shows the strong promise of a full data combination. 

Comparing the different combinations with each other we see that there is a preference to the ordering of PTA additions. Starting with NANOGrav data first in the process results in the highest SNRs, and adding EPTA+InPTA data before PPTA results in higher SNRs. These preferences are consistent with each PTA's individual SNR.

\section{Conclusions}

We performed a comparison of the noise models used and the properties of the GWB recovered in the most recent data-sets of \citetalias{epta+inpta}, \citetalias{ng15}, and \citetalias{pptadr3}.
We found that a majority of the noise parameters were consistent between the three different data sets.
Where there were significant ($>3\sigma$) differences, they could be attributed to different time spans and cadences as well as the lack of frequency coverage in individual PTAs.
These tensions also reduced significantly when standardized noise models were adopted.
We also calculated and compared the relative sensitivity of the three data sets using sensitivity curves, showing that the different levels of reported evidence for the GWB were consistent with the sensitivities of each of these data sets. 

Despite the different noise models used by each PTA, the GWB posterior distributions for all three PTA data sets were consistent within $1\sigma$ as calculated by the tension metric.
The GWB spectra measured from all three PTA data sets were consistent with a single, ``joint'' power law spectrum, with no evidence for deviations from this power law spectral template.

Finally, we extended each of the three data sets using the factorized likelihood approach, to achieve a pseudo-IPTA analysis. The amplitudes of the CURN process estimated from a global posterior using different permutations of PTA extensions were found to be broadly consistent with each other as well as with those reported by the individual PTAs. The addition of pulsars to any PTA also resulted in an increase in the measured HD SNR.

The members of the IPTA, along with the MeerKAT PTA \citep{mpta_dr1}, are currently in  the process of combining their most recent data sets, which will become IPTA-DR3.
The comparisons presented here motivate improvements and best practices to be adopted in a unified analysis for the ongoing data combination by the IPTA.
Based on our work we believe that choosing the right noise model for each pulsar through Bayesian model section will be an important step for future IPTA analyses.
In particular great care must be taken with the pulsars J0030+0451, J0613$-$0200, J1022+1001, J1640+2224, J1713+0747, J1744$-$1134 and J2124$-$3358 where discrepancies between PTAs persisted when using the standardized noise model.
While the results presented here adopted a much simplified approach as opposed to a true combination, these already hint at an enhancement in the significance of GWB detection in the full DR3 over those reported by \citetalias{epta+inpta}, \citetalias{ng15}, and \citetalias{pptadr3} individually.

\section*{Acknowledgments}
\textit{Author contributions.}
An alphabetical-order author list was used for this paper to recognize the vast extent of the time and effort by many people that went into this project.
All authors contributed to the activities of the IPTA collaboration leading to the work presented here, and reviewed the text and figures prior to the paper’s submission.
This project was organized as part of the Gravitational Wave Analysis Working Group of the IPTA by PTB, ACh, and NSP.

SDan and PTB lead the GW comparisons with contributions from LD, SDe, and AG.
ACh lead the noise comparisons with contributions from BBL, SDan, CD, FK, DDeb, and CMFM.
GS lead the GW sensitivity analysis with contributions from KGr, NKP, JSH, and KEW.
KAG, LSch, and NSP performed the factorized likelihood and optimal statistic analyses.
WGL performed the free spectral refitting.

PTB, ACh, NSP, SDan, KAG, LSch, WGL, BCJ, and GS wrote the text and generated figures.

\textit{Acknowledgments.}
% main collaboration stuff
The European Pulsar Timing Array (EPTA) is a collaboration between European and partner institutes, namely ASTRON (NL), INAF/Osservatorio di Cagliari (IT), Max-Planck-Institut für Radioastronomie (GER), Nançay/Paris Observatory (FRA), the University of Manchester (UK), the University of Birmingham (UK), the University of East Anglia (UK), the University of Bielefeld (GER), the University of Paris (FRA), the University of Milan-Bicocca (IT), the Foundation for Research and Technology, Hellas (GR), and Peking University (CHN), with the aim to provide high-precision pulsar timing to work towards the direct detection of low-frequency gravitational waves.
The Indian Pulsar Timing Array (InPTA) is an Indo-Japanese collaboration that routinely employs TIFR’s upgraded Giant Metrewave Radio Telescope for monitoring a set of IPTA pulsars.
The NANOGrav collaboration receives support from National Science Foundation (NSF) Physics Frontiers Center award Nos.~1430284 and 2020265, the Gordon and Betty Moore Foundation, NSF AccelNet award No.~2114721, an NSERC Discovery Grant, and CIFAR.
Part of this research was undertaken as part of the Australian Research Council (ARC) Centre of Excellence for Gravitational Wave Discovery (OzGrav) under grant CE170100004.

% telescopes
An Advanced Grant of the European Research Council allowed to implement the Large European Array for Pulsars (LEAP) under Grant Agreement No.~227947 (PI M.~Kramer).
Part of this work is based on observations with the 100-m telescope of the Max-Planck-Institut für Radioastronomie (MPIfR) at Effelsberg in Germany.
Pulsar research at the Jodrell Bank Centre for Astrophysics and the observations using the Lovell Telescope are supported by a Consolidated Grant (ST/T000414/1) from the UK’s Science and Technology Facilities Council (STFC).
The Nançay radio Observatory is operated by the Paris Observatory, associated with the French Centre National de la Recherche Scientifique (CNRS), and partially supported by the Region Centre in France.
We acknowledge financial support from ``Programme National de Cosmologie and Galaxies'' (PNCG), and ``Programme National Hautes Energies'' (PNHE) funded by CNRS/INSU-IN2P3-INP, CEA and CNES, France.
We acknowledge financial support from Agence Nationale de la Recherche (ANR-18-CE31-0015), France.
The Westerbork Synthesis Radio Telescope is operated by the Netherlands Institute for Radio Astronomy (ASTRON) with support from the Netherlands Foundation for Scientific Research (NWO).
The Sardinia Radio Telescope (SRT) is funded by the Department of University and Research (MIUR), the Italian Space Agency (ASI), and the Autonomous Region of Sardinia (RAS) and is operated as a National Facility by the National Institute for Astrophysics (INAF).
The Arecibo Observatory is a facility of the NSF operated under cooperative agreement (AST-1744119) by the University of Central Florida (UCF) in alliance with Universidad Ana G.~M\'endez (UAGM) and Yang Enterprises (YEI), Inc.
The Green Bank Observatory is a facility of the NSF operated under cooperative agreement by Associated Universities, Inc.
The National Radio Astronomy Observatory is a facility of the NSF operated under cooperative agreement by Associated Universities, Inc.
Murriyang, the Parkes 64 m radio telescope is part of the Australia Telescope National Facility (\url{https://ror.org/05qajvd42}), which is funded by the Australian Government for operation as a National Facility managed by CSIRO. We acknowledge the Wiradjuri People as the Traditional Owners of the Observatory site.

% computing
This work made use of the OzSTAR national facility at Swinburne University of Technology. OzSTAR is funded by Swinburne University of Technology and the National Collaborative Research Infrastructure Strategy (NCRIS).
This work was conducted in part using the resources of the Advanced Computing Center for Research and Education (ACCRE) at Vanderbilt University, Nashville, TN.
This work used resources of the IN2P3 Computing Center (CC-IN2P3 - Lyon/Villeurbanne - France) funded by the Centre National de la Recherche Scientifique.
%We acknowledge the National Supercomputing Mission (NSM) for providing computing resources of `PARAM Ganga' at IIT Roorkee,
%and `PARAM Seva' at IIT Hyderabad, which is implemented by C-DAC and supported by the Ministry of Electronics and Information Technology (MeitY) and Department of Science and Technology (DST), Government of India.
%LSp acknowledges the use of the HPC system Cobra at the Max Planck Computing and Data Facility.
%This work was conducted using the Thorny Flat HPC Cluster at West Virginia University (WVU), which is funded in part by National Science Foundation (NSF) Major Research Instrumentation Program (MRI) Award No.~1726534, and West Virginia University.
%We also thank CSIRO Information Management and Technology High Performance Computing group for access and support with the petrichor cluster.

% EPTA
The work is supported by the National SKA programme of China (2020SKA0120100), Max-Planck Partner Group, NSFC 11690024, CAS Cultivation Project for FAST Scientific.
This work is also supported as part of the ``LEGACY'' MPG-CAS collaboration on low-frequency gravitational wave astronomy.
JA acknowledges support from the European Commission (Grant Agreement No.~101094354).
JA and SChan were partially supported by the Stavros Niarchos Foundation (SNF) and the Hellenic Foundation for Research and Innovation (HFRI) under the 2nd Call of the ``Science and Society – Action Always strive for excellence – Theodoros Papazoglou'' (Project No.~01431).
ACh acknowledges support from the Paris Île-de-France Region.
ACh, AF, ASe, ASa, EB, DI, GMS, MBo acknowledge financial support provided under the European Union’s H2020 ERC Consolidator Grant “Binary Massive Black Hole Astrophysics” (B Massive, Grant Agreement: 818691).
GD, KLi, RK and MKr acknowledge support from European Research Council (ERC) Synergy Grant “BlackHoleCam”, Grant Agreement No.~610058.
ICN is supported by the STFC doctoral training grant ST/T506291/1.
AV and PRB are supported by the UK’s Science and Technology Facilities Council (STFC; grant ST/W000946/1).
AV also acknowledges the support of the Royal Society and Wolfson Foundation.
NKP is funded by the Deutsche Forschungsgemeinschaft (DFG, German Research Foundation) – Projektnummer PO 2758/1–1, through the Walter–Benjamin programme.
ASa thanks the Alexander von Humboldt foundation in Germany for a Humboldt fellowship for postdoctoral researchers.
APo, DP and MBu acknowledge support from the research grant ``iPeska'' (P.I.~Andrea Possenti) funded under the INAF national call Prin-SKA/CTA approved with the Presidential Decree 70/2016 (Italy).
RNC acknowledges financial support from the Special Account for Research Funds of the Hellenic Open University (ELKE-HOU) under the research programme ``GRAVPUL'' (grant agreement 319/10-10-2022).
EvdW, CGB and GHJ acknowledge support from the Dutch National Science Agenda, NWA Startimpuls – 400.17.608.
BG is supported by the Italian Ministry of Education, University and Research within the PRIN 2017 Research Program Framework, No.~2017SYRTCN.

% INPTA
BCJ, YG, YM, SDan, AG and PR acknowledge the support of the Department of Atomic Energy, Government of India, under Project Identification \# RTI 4002.
BCJ, YG and YM acknowledge support of the Department of Atomic Energy, Government of India, under project No.~12-R\&D-TFR-5.02-0700, while SD, AG and PR acknowledge support of the Department of Atomic Energy, Government of India, under project No.~12-R\&D-TFR-5.02-0200.
KT is partially supported by JSPS KAKENHI Grant Nos.~20H00180, 21H01130, and 21H04467, Bilateral Joint Research Projects of JSPS, and the ISM Cooperative Research Program (2021-ISMCRP-2017).
AKP is supported by CSIR fellowship Grant No.~09/0079(15784)/2022-EMR-I.
SH is supported by JSPS KAKENHI Grant No.~20J20509.
KN is supported by the Birla Institute of Technology \& Science Institute fellowship.
AmS is supported by CSIR fellowship Grant No.~09/1001(12656)/2021-EMR-I and T-641 (DST-ICPS).
TK is partially supported by the JSPS Overseas Challenge Program for Young Researchers.
DDeb acknowledges the support from the Department of Atomic Energy, Government of India through Apex Project - Advance Research and Education in Mathematical Sciences at IMSc.

% NANOGrav
LB acknowledges support from the National Science Foundation under award AST-1909933 and from the Research Corporation for Science Advancement under Cottrell Scholar Award No.~27553.
SB gratefully acknowledges the support of a Sloan Fellowship, and the support of NSF under award \#1815664.
The work of RB, RC, DD, NLa, XS, JPS, and JT is partly supported by the George and Hannah Bolinger Memorial Fund in the College of Science at Oregon State University.
MC, PP, and SRT acknowledge support from NSF AST-2007993.
MC and NSP were supported by the Vanderbilt Initiative in Data Intensive Astrophysics (VIDA) Fellowship.
KCh, ADJ, and MV acknowledge support from the Caltech and Jet Propulsion Laboratory President’s and Director’s Research and Development Fund.
KCh and ADJ acknowledge support from the Sloan Foundation.
Support for this work was provided by the NSF through the Grote Reber Fellowship Program administered by Associated Universities, Inc./National Radio Astronomy Observatory.
Support for HTC is provided by NASA through the NASA Hubble Fellowship Program grant \#HST-HF2-51453.001 awarded by the Space Telescope Science Institute, which is operated by the Association of Universities for Research in Astronomy, Inc., for NASA, under contract NAS5-26555.
KCr is supported by a UBC Four Year Fellowship (6456).
MED acknowledges support from the Naval Research Laboratory by NASA under contract S-15633Y.
TD and MTL are supported by an NSF Astronomy and Astrophysics Grant (AAG) award No.~2009468.
ECF is supported by NASA under award No.~80GSFC21M0002.
GEF, SCS, and SJV are supported by NSF award PHY2011772.
KAG and SRT acknowledge support from an NSF CAREER award \#2146016.
The Flatiron Institute is supported by the Simons Foundation.
SH is supported by the National Science Foundation Graduate Research Fellowship under Grant No.~DGE-1745301.
NLa acknowledges the support from Larry W.~Martin and Joyce B.~O’Neill Endowed Fellowship in the College of Science at Oregon State University.
Part of this research was carried out at the Jet Propulsion Laboratory, California Institute of Technology, under a contract with the National Aeronautics and Space Administration (80NM0018D0004).
DRL and MAMc are supported by NSF \#1458952.
MAMc is supported by NSF \#2009425.
CMFM was supported in part by the National Science Foundation under Grants No.~NSF PHY-1748958 and AST-2106552.
AMi is supported by the Deutsche Forschungsgemeinschaft under Germany’s Excellence Strategy - EXC 2121 Quantum Universe - 390833306.
PN acknowledges support from the BHI, funded by grants from the John Templeton Foundation and the Gordon and Betty Moore Foundation.
The Dunlap Institute is funded by an endowment established by the David Dunlap family and the University of Toronto.
KDO was supported in part by NSF Grant No.~2207267.
TTP acknowledges support from the Extragalactic Astrophysics Research Group at E\"otv\"os Lor\'and University, funded by the E\"otv\"os Lor\'and Research Network (ELKH), which was used during the development of this research.
SMR and IHS are CIFAR Fellows.
Portions of this work performed at NRL were supported by ONR 6.1 basic research funding.
JDR also acknowledges support from start-up funds from Texas Tech University.
JS is supported by an NSF Astronomy and Astrophysics Postdoctoral Fellowship under award AST-2202388, and acknowledges previous support by the NSF under award 1847938.
CU acknowledges support from BGU (Kreitman fellowship), and the Council for Higher Education and Israel Academy of Sciences and Humanities (Excellence fellowship).
CAW acknowledges support from CIERA, the Adler Planetarium, and the Brinson Foundation through a CIERA-Adler postdoctoral fellowship.
OY is supported by the National Science Foundation Graduate Research Fellowship under Grant No.~DGE2139292.

% PPTA
RMS acknowledges support through ARC Future Fellowship FT190100155.
SDa is the recipient of an Australian Research Council Discovery Early Career Award (DE210101738) funded by the Australian Government.
YL acknowledges support of the Simons Investigator Grant 827103.

\bibliography{bib}

\allauthors
\end{document}